\documentclass[12pt,preprint]{aastex}

\newcommand{\etal}{{\em et al.\ }}

\def\ion[#1 #2]{#1\,{\sc #2}}
\def\ergs[#1]{#1 {ergs}~{cm$^{-2}$}\,{s$^{-1}$}\,{sr$^{-1}$}}
\def\dens[#1]{10$^{#1}$\hskip 1.5pt{cm$^{-3}$}}
\def\densr[#1 #2]{10$^{#1}$\hskip 1pt{--}\hskip .5pt{10$^{#2}$}\hskip 1.5pt{cm$^{-3}$}}
\def\fl[#1 #2]{{#1}$\pm${#2}}
\def\orb[#1 #2]{{$#1^{#2}$}}
\def\ls[#1 #2]{{$^{#1}${#2}}}
\def\tm[#1 #2 #3]{{$^{#1}${#2}$_{#3}$}}
\def\FLASH                 {{\sc flash}}
\def\PARAMESH              {{\sc paramesh}}

\begin{document}

\title{Post-flare UV light curves explained with thermal instability of loop plasma}

\author{F. Reale\altaffilmark{1,2}, E. Landi,\altaffilmark{3}, S. Orlando\altaffilmark{2}}

\altaffiltext{1}{Dipartimento di Fisica, Universit\`a degli Studi di Palermo, Piazza del Parlamento 1,
90134, Italy}
\altaffiltext{2}{INAF-Osservatorio Astronomico di Palermo, Piazza del Parlamento 1, 90134 Palermo, Italy}
\altaffiltext{3}{Department of Atmospheric, Oceanic and Space Sciences, University of Michigan, Ann Arbor, MI 48109}

\begin{abstract}
In the present work we study the C8 flare occurred on September 26, 2000 at 19:49~UT
and observed by the SOHO/SUMER spectrometer from the beginning of the impulsive phase 
to well beyond the disappearance in the X-rays. The emission first decayed progressively 
through equilibrium states until the plasma reached 2-3~MK. Then, a series of cooler 
lines, i.e. \ion[Ca x] , \ion[Ca vii], \ion[Ne vi], \ion[O iv] and \ion[Si iii] (formed in 
the temperature range $\log T=4.3-6.3$ under equilibrium conditions), are emitted at the 
same time and all evolve in a similar way. Here we show that the simultaneous emission of 
lines with such a different formation temperature is due to thermal instability occurring
in the flaring plasma as soon as it has cooled below $\sim 2$ MK. We can qualitatively
reproduce the relative start time of the light curves of each line in the correct order
with a simple (and standard) model of a single flaring loop. The agreement with the 
observed light curves is greatly improved, and a slower 
evolution of the line emission is predicted, if we assume that the model loop consists 
of an ensemble of subloops or strands heated at slightly different times. Our analysis 
can be useful for flare observations with SDO/EVE.
\end{abstract}


\section{Introduction}

Flares are one of the most important components of solar activity and
result in sudden releases of magnetic energy and in massive magnetic
field restructuring in the host active regions. They are made of two
distinct phases: a very short impulsive phase - when magnetic energy is 
suddenly released and plasma is rapidly heated to temperatures that may
exceed 15~MK - and a long decay phase, where the plasma cools back to 
lower temperatures. Most of flare observations have been carried out 
using crystal spectrometers 
observing below 20~\AA\ or narrow-band imagers with bandpasses optimized 
at wavelengths below 50~\AA, where the emission from multimillion degrees 
plasmas is best observed. X-ray spectral observations of the decay phase 
of flares have invarably shown a slow decrease of the plasma temperature, 
down to a few million degrees. A large body of literature has been devoted 
to study the properties of such decaying plasmas. When the flare plasma 
temperature decreases below a few million degrees, it fades from X-ray 
instruments and cannot be studied anymore with X-ray spectroscopic diagnostic 
techniques. As a consequence, no diagnostic studies could be made of
the evolution of the post-flare plasmas in the last phases of cooling.
EUV imagers, such as SOHO/EIT, TRACE, STEREO/EUVI and SDO/AIA, sample quiescent 
coronal temperatures of $\approx 1-3$~MK and thus can extend decay phase 
flare observations down to lower temperatures. However, since they rely 
on broad-band filters, they can only provide limited diagnostic results.
SUMER, observing in the 500-1600~\AA\ UV range, provides a golden opportunity 
to monitor the behavior of decaying plasma down to chromospheric temperatures, 
thanks to the multitude of coronal, transition region and chromospheric lines 
that populate its spectral range. During its operational life, SUMER 
observed a number of C and M flares, 
and some of its observations lasted enough to let the post-flare plasma cover 
the entire evolution from the impulsive to the very end of the decay phase. 
However, the spectroscopic diagnostic potential of SUMER has never been 
exploited to try to study the evolution of the post-flare plasma below 2~MK. 
The aim of the present work is to carry out such a study.

Recently, Feldman \etal (2003) reported on SUMER observations of the evolution
a C8 flare from the impulsive phase down to below 1~MK. The advantage of their 
observations was that they covered the 1097-1119~\AA\ spectral range, which
includes spectral lines formed at all temperatures from 0.01~MK to 15~MK.
The observation lasted enough to allow them to follow the evolution of the
flare plasma from the impulsive phase to well beyond the lower temperature 
limit imposed by the use of X-ray instrumentation. They found that the 
plasma continued to cool through quasi-equilibrium states until 
it reached $\approx$2-3~MK. Below that threshold, Feldman \etal (2003) 
reported a dramatic change in the emission, with the surprising, sudden 
appearance at the same time of lines from ions formed between 0.01~MK and 
1~MK that were not expected to be observed simultaneously in a plasma in 
ionization equilibrium. Their observations suggested that the plasma was
undergoing thermal non-equilibrium. Feldman \etal (2003) did not investigate 
this final feature of their observation, and limited themselves to reporting 
the light curves of the cooling flare plasma. The aim of the present work 
is to revisit the observations of Feldman \etal (2003) and investigate in detail 
the scenario of thermal non-equilibrium in the final phase of the flare decay,
using a state-of-the-art time-dependent loop hydrodynamic code which includes 
non equilibrium ionization.  In particular, we focus on predicting the 
light curves of all lines emitted in the late phase of flare decay, and on comparing 
them to the SUMER observations. We will show that thermal non-equilibrium, 
i.e. a catastrophic cooling of a coronal plasma, develops naturally in a standard 
flaring loop simulation late in the decay in agreement with observations, and that 
it is able to 
explain well the sequence and timing of appearance of the observed lines. 

This observation represents an excellent testing ground of the 
presence of thermal non-equilibrium in the Sun, and it is expected to be 
of significant value to study several phenomena other than flares. 
In fact, thermal non-equilibrium may produce plasma flows, the formation 
of condensations and the loss of connection between electron temperature 
and the ionization status of a plasma. Thermal non-equilibrium has been suggested to 
be responsible for the formation of prominences from coronal loops (e.g. Antiochos 
1980, Karpen \etal 2006 and references therein) and intensity variations traveling 
along the loop structure (M\"uller \etal 2005) observed with the \ion[He ii] SOHO/EIT 
band (De Groof \etal 2004). Also, it has been recently suggested as a possible 
explanation of the main observational properties of coronal loops, including 
lifetime, temperature and density profiles, and filamentary structure (Klimchuk 
\etal 2010).

In addition, observations of the decay phase of flares have been recently made with 
the SDO/EVE instrument (Woods \etal 2010), which observes the emission of the 
entire Sun between 1 and 1060~\AA\ at a reduced resolution of 1~\AA\ (60-1060~\AA\ 
range) and 10~\AA\ (1-60~\AA\ range). EVE flare decays showed qualitatively the 
same features described by Feldman \etal (2003), with a series of intensity 
enhancements of lines from progressively colder ions, down to around 1~MK. 
Since EVE observes continuously the entire solar disk, flare observations are
being made routinely, although at a reduced spectral resolution. Thus, the 
high-resolution results of the present work can be used as a guideline for 
the interpretation of EVE flare observations.

The observations we use are summarized in Section~\ref{observations}, where 
we also refine the measurement of the light curves of all available lines and 
improve on the electron density measurements over the results of Feldman \etal 
(2003). The hydrodynamic code and numerical setup are described in 
Section~\ref{model}, and the comparison between model predictions and 
observations is reported in Section~\ref{results}. Section~\ref{discussion}
discusses the results and summarizes this work.

\section{The SUMER picture of a C8 flare}
\label{observations}

\subsection{Observations}

The observations studied in this work are fully described in Feldman 
\etal (2003); here we only summarize their main features. The data 
set consists of a series of spectra observed by the SUMER spectrometer 
on board SOHO (Wilhelm \etal 1995) while its 4\arcsec$\times$300\arcsec\
slit was held fixed with the center pointed at (-980\arcsec,-250\arcsec).
Thus, the entire field of view of the SUMER slit was placed outside 
the solar limb, above an active region. The 1097-1119~\AA\ wavelength 
range was transmitted to the ground after being observed with an 
exposure time of 162.5~s. The choice of such a limited wavelength 
range was dictated by the need of a relatively fast cadence, but 
it allowed us to include spectral lines from ions formed at all 
temperatures of the solar corona. Table~\ref{lines} lists them 
along with the temperature of maximum ion abundance from Bryans 
\etal (2009); free-free continuum radiation was also observed 
during the flare.

The observations were carried out on September 26, 2000 during 
the peak of solar cycle 23. They started at 11:41~UT and ended 
at 23:07~UT. During the observing run a C8 flare erupted from 
the active region and was observed by SUMER. X-ray data from 
GOES show that it started at 19:49~UT and by 20:20~UT it had 
faded in the X-ray background. Despite the disappearance from
GOES X-ray data, the flare plasma kept evolving: thanks to 
the wide range of formation temperature of the lines in the 
1097-1119~\AA\ range, SUMER was able to observe the flare for 
a much wider time interval, spanning from the very beginning 
of the impulsive phase to much beyond the end of the GOES X-ray 
flare. 

\subsection{Flare emission evolution}

\begin{figure}
\includegraphics[width=15.0cm,height=17.0cm,angle=90]{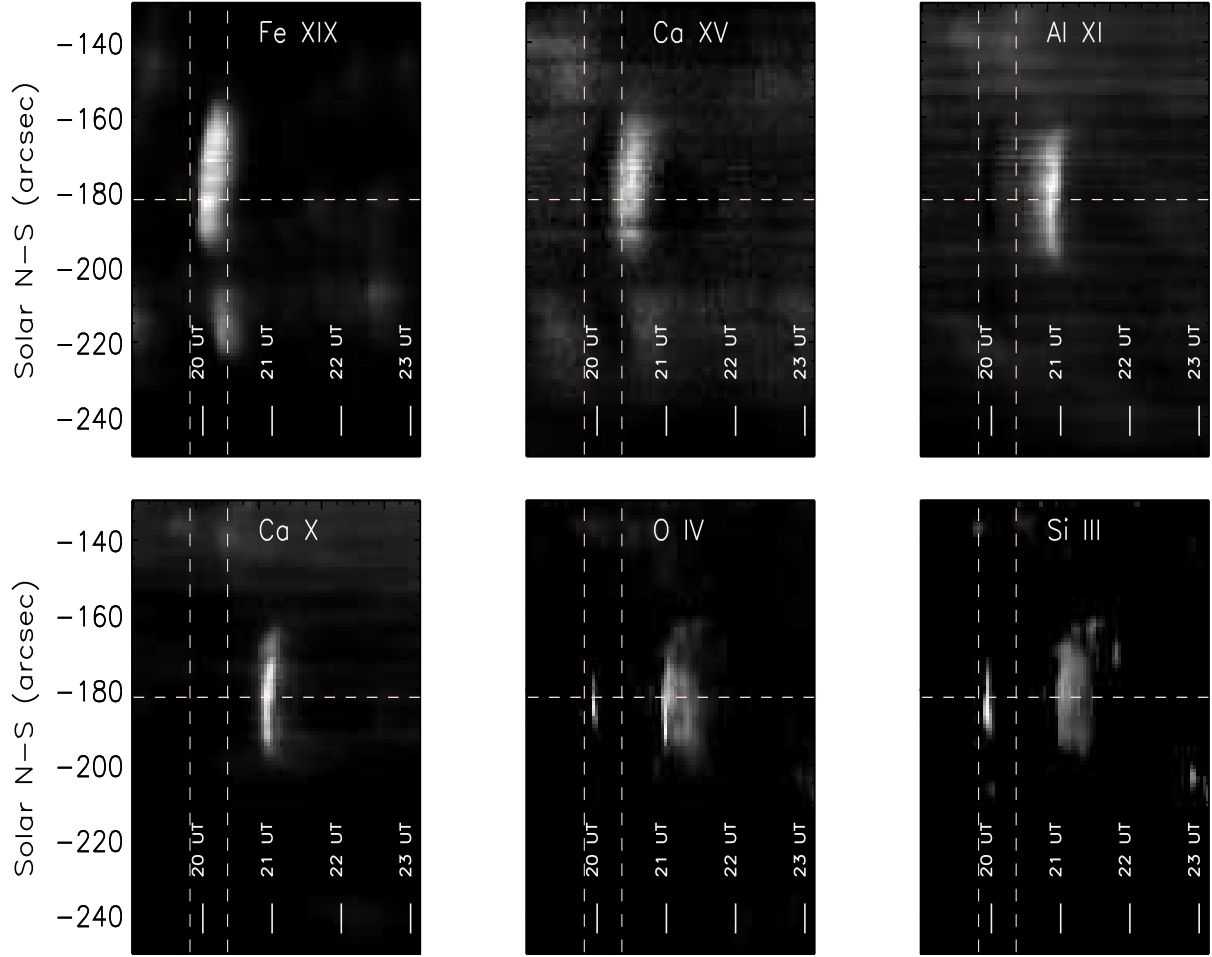}
\caption{\label{time_int} Time-intensity plot of selected ions observed by
SUMER. The X-axis is time, the Y-axis is the position along the slit in
heliocentric coordinates. The two vertical lines in each panel represent 
the beginning and the end of the GOES flare. The horizontal line indicates
the pixel of the SUMER slit chosen for the analysis. The \ion[Si iii] 
1113.23~\AA\ and \ion[O iv] 554.08~\AA\ lines are displayed here.}
\end{figure}

Figure~\ref{time_int} illustrates the behavior of the plasma. Each panel is 
built in two steps.  First, the intensity of a line is measured at all pixels 
along the slit for each of the 162.5~s-long exposures. Then, these 1-D intensity 
profiles are placed one next to the other to create a 2-D image, where the 
X-axis represents time, and the Y-axis represents the coordinate along the 
slit. GOES start and end times of the flare are added to each panel of
Figure~\ref{time_int} as vertical dashed lines. Figure~\ref{time_int} 
captures the three main features of the flare:

\begin{enumerate}

\item The impulsive phase shows a sudden, very localized enhancement 
of \ion[Si iii] and \ion[O iv] line emission, followed after a few minutes 
by a large and long-lasting brightening of \ion[Fe xix] line emission;

\item The decay phase lasts far longer than observed by GOES, indicating 
that X-ray instruments only allow to observe a limited portion of the
decay phase of the flare;

\item The final phase of the flare can be observed simultaneously with
\ion[O iv] and \ion[Si iii] (as well as \ion[Ne vi] and \ion[Ca vii],
not shown in the figure).

\end{enumerate}

The presence of two distinct loop sources as responsible of different 
phases of a solar flare was remarked by Reale \etal (2004), who modeled 
a flare in Proxima Centauri observed with XMM with two different 
populations of loops. In a similar way, the 
initial \ion[Si iii] and \ion[O iv] pulses observed by SUMER in the 26 
Septemper 2000 flare may be due to a different kind of loops than the 
rest of the decay phase observed with all the other ions, and thus can 
not be modeled together with the latter in a single-loop framework.

\begin{figure}
\includegraphics[width=15.0cm,height=17.0cm,angle=90]{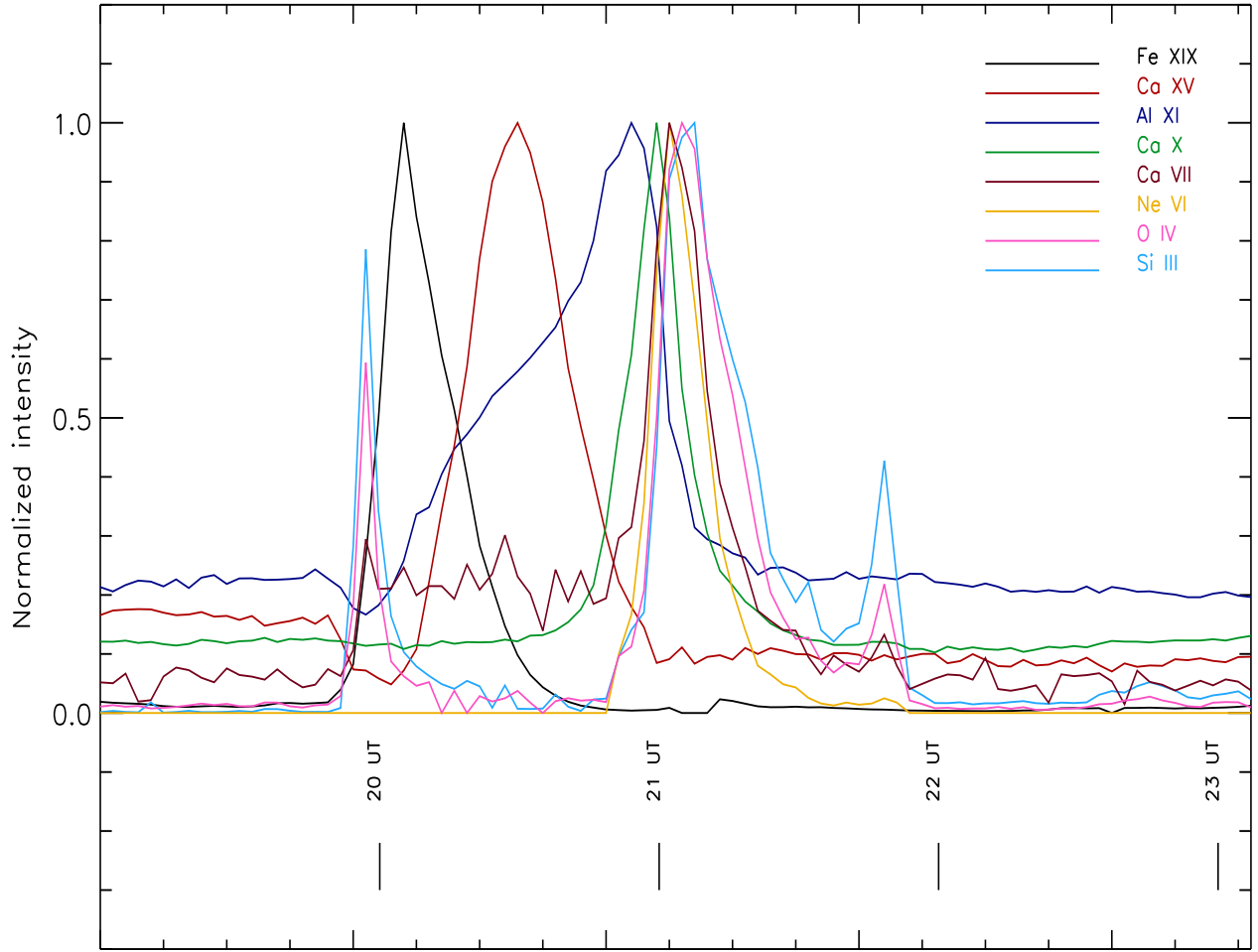}
\caption{\label{light_curves} Normalized intensities of several lines
measured as a function of time from the selected pixel along the SUMER 
slit. The \ion[Si iii] 1113.23~\AA\ and \ion[O iv] 554.08~\AA\ 
lines are displayed here.}
\end{figure}

Feldman \etal (2003) found that the plasma emitting during the decay 
phase of the flare between 20~UT and 21~UT was close to isothermal, 
and its temperature decreased exponentially with time from 7.9~MK to 
2.5~MK. Then, they reported the simultaneous emission of several ions 
whose temperature of formation under ionization equilibrium spanned 
more than one order of magnitude. The light curves of these cooler 
ions suggest that they are emitted by the cooling post-flare plasma.
In fact, Figure~\ref{time_int} shows that the intensity enhancements 
of \ion[Ca x], \ion[O iv] and \ion[Si iii] all overlap, and they prolong
in time those of \ion[Fe xix], \ion[Ca xv] and \ion[Al xi] studied
by Feldman \etal (2003). To provide quantitative measurements of the 
relationship among the light curves of all these ions we chose to 
concentrate on a single pixel along the SUMER slit, shown in 
Figure~\ref{time_int} by the horizontal dashed line. This pixel 
corresponds to ``position~38'' in Feldman \etal (2003). We measured
the intensities of all lines in this position as a function of time, 
and displayed their normalized values in Figure~\ref{light_curves}. 
The impulsive phase of the flare is visible by a sudden brightening 
in \ion[Si iii] and \ion[O iv] right before 20~UT, and then the long 
decay phase studied by Feldman \etal (2003) is apparent from the peak 
of the \ion[Fe xix] line emission to that of \ion[Al xi]. 

The surprising feature of Figure~\ref{light_curves} is the sudden 
appearance (at around 21~UT) and co-existence of the emission from 
\ion[Ca x] (formed in equilibrium at $\log T=5.7-6.3$), \ion[Ca vii] 
($\log T=5.4-5.9$), \ion[Ne vi] ($\log T=5.4-5.8$), \ion[O iv] ($\log 
T=4.9-5.5$) and \ion[Si iii] ($\log T=4.3-5.0$). These ions are emitted 
by a previously cooling, near-isothermal plasma undergoing a series of 
equilibrium states, and yet cannot be emitted simultaneously 
by an isothermal plasma in ionization equilibrium as their temperature 
ranges of formation do not overlap. The most likely explanation to the 
presence of the emission from these ions is that the cooling post-flare 
plasma undergoes a thermal instability as it reaches $\approx 2$~MK.

The modeling effort of the present work is aimed at reproducing the light 
curves of the thermal instability after 21~UT. We are particularly interested 
at studying the succession of rise and fall of the emission in the colder 
ions.

\subsection{Plasma diagnostics}
\label{diagnostics_results}

The plasma temperature was estimated by Feldman \etal (2003) assuming
that the plasma evolved through quasi-equilibrium states from right after
the impulsive phase (around 20:05~UT, approximately 15 minutes after the 
GOES beginning of the flare) down to 20:45~UT, so that the shape of the 
\ion[Fe xix] and \ion[Ca xv] light curves could be fitted using the 
contribution functions of the two lines. The $\log T$ of the plasma 
was found to evolve linearly with time from an initial $\log T=6.9$ 
at 20:05~UT to $\log T=6.40$ at 20:45~UT. The free-free emission was 
observed until 21:05~UT and it allowed to determine that the plasma 
composition was similar to the standard composition of the solar corona; 
this means that the plasma in the decay phase of the flare had a local, 
coronal origin rather than being evaporated from the chromosphere.

The plasma density was determined by Feldman \etal (2003) by using the 
value of the total emission measure they derived from the free-free 
radiation and an assumed value of the total length of the 
flaring plasma along the line of sight. However, there are two ions 
in the 1097-1119~\AA\ range that emit more than one line: \ion[Si iii] 
and \ion[O iv]. The former provides a ratio which is
density sensitive in the $\log N_e=10-12$ range ($N_e$ in cm$^{-3}$), 
so that we can attempt a crude estimate using the emissivities from 
the CHIANTI database (Dere \etal 1995, 2009). \ion[Si iii] emission 
is observed at the very beginning of the flare, during the impulsive 
phase, and late in the decay phase. At the beginning of the event 
the intensity ratio is 0.60$\pm$0.06, corresponding to $\log N_e > 11.8$ 
while at the end of 
the decay phase the ratio is $0.43\pm0.03$, with little temporal evolution, 
indicating a density of $\log N_e =10.5\pm0.3$. The density estimate 
made by Feldman \etal (2003) during the impulsive phase 
is consistent with the value obtained with the line intensity ratio 
only if we assume that the actual volume filled by the plasma is very 
small, indicating a filling factor smaller than unity. During the decay 
phase, on the contrary, the two estimates agree with each other if we 
assume a unity filling factor and a length of the line of sight $H$ in the 
1\arcsec-10\arcsec\ range. On Sept. 26, 2000 1\arcsec\ at SOHO's 
distance corresponded to $\approx$720~km at the Sun, so that
$7.2\times 10^7 \leq H \leq 7.2\times 10^8$~cm.

\section{Hydrodynamic modeling and numerical setup}
\label{model}

We aim at explaining the evolution of the observed emission through the
hydrodynamic modeling of a flare-heated plasma confined in a coronal 
magnetic flux tube as in Reale \& Orlando (2008). We consider 
semicircular loops with half-length in the range 
$1.5 \times 10^9 < L < 5 \times 10^9$ cm, typical of active region loops, 
and constant cross-section area all along the loops. The loops lie on a 
plane vertical to the solar surface. It is assumed that the loop plasma 
moves and transports energy only along the magnetic field lines. The 
plasma evolution can then be described by one-dimensional (1-D) 
hydrodynamics (e.g., Peres \etal 1982):

\begin{equation}
\frac{\partial \rho}{\partial t} + \frac{\partial \rho u}{\partial s} = 0
\label{eq1}
\end{equation}

\begin{equation}
\frac{\partial \rho u}{\partial t} + \frac{\partial (P+\rho
u^2)}{\partial s} = \rho g
\end{equation}

\begin{equation}
\frac{\partial \rho E}{\partial t} +\frac{\partial (\rho E+P) u}
{\partial s} = \rho u g - \frac{\partial q}{\partial s} + Q(s, t) -n_e
n_H \Lambda(T)
\end{equation}

\[
\mbox{where \hspace{0.5cm}} E = \epsilon +\frac{1}{2} u^2
\]

\noindent
is the total gas energy (internal energy, $\epsilon$, and kinetic
energy), $t$ the time, $s$ the coordinate along the loop, $\rho =
\mu m_H n_{\rm H}$ the mass density, $\mu = 1.26$ the mean atomic
mass (assuming solar abundances), $m_H$ the mass of the hydrogen
atom, $n_{\rm H}$ the hydrogen number density, $n_{\rm e}$ the
electron number density, $u$ the plasma flow velocity, $P$ the pressure,
$g$ the component of gravity parallel to the field lines, $T$ the
temperature, $q$ the conductive flux, $Q(s,t)$ a function describing
the transient input heating, $\Lambda(T)$ the radiative losses per
unit emission measure (Rosner et al. 1978). The radiative losses
are assumed from plasma in ionization equilibrium and are set
to zero for $T < 20000$ K, to ensure the
detailed energy balance in the chromosphere. Tests
have shown that the simulation results do not change considerably because of
local and temporary changes of the radiative losses function,
as those due by transient deviations from equilibrium of ionization (Reale \&
Orlando 2008).  We use the ideal gas law, $P=(\gamma-1) \rho \epsilon$,
where $\gamma=5/3$ is the ratio of specific heats.

The flare evolution may be faster than ionization and recombination
timescales, so that the ionization fraction of the emitting ions may not
have enough time to adjust to the rapidly changing plasma temperature (e.g. 
Golub \etal 1989, Orlando \etal 1999). 
This effect may be important even in the late phases of the flare. 
Therefore in our modeling we also compute the ionization fractions of the 
most important elements (namely He, C, N, O, Ne, Mg, Si, S, Ar, Ca, Fe, Ni), 
by solving synchronously but independently the continuity equations for each 
ion species:

\begin{equation}
\frac{\partial n_i^Z}{\partial t} + \nabla \cdot n_i^Z \mbox{\bf v} =
R_i^Z ~~~~~~~~\begin{array}{l}(Z = 1, ..., N_{elem})\\\\
(i = 1, ..., N_{ion}^Z) \end{array}
\label{eq4}
\end{equation}

\noindent
\[
\mbox{where}~~~R_i^Z = n_e [n_{i+1}^Z\alpha_{i+1}^Z + n_{i-1}^Z
S_{i-1}^Z - n_i^Z(\alpha_i^Z+S_i^Z)]~,
\]

\noindent
$n_{i}^Z$ is the number density of the $i$-th ion of the element $Z$,
$N_{elem}$ the number of elements, $N_{ion}^Z$ the number of ionization
states of element $Z$, $\alpha_i^Z$ the collisional and dielectronic
recombination coefficients, and $S_i^Z$ the collisional ionization
coefficients (Summers 1974).

As discussed in Reale \& Orlando (2008), the heat conduction may become 
``flux limited'' during flares (e.g. Brown \etal 1979), and our model 
allows for a smooth transition between the classical and saturated 
conduction fluxes (Dalton \& Balbus 1993):

\begin{equation}
q = \left(\frac{1}{q_{\rm spi}}+\frac{1}{q_{\rm sat}}\right)^{-1}~,
\end{equation}

\noindent
where $q_{\rm spi}=-\kappa(T)\nabla T$ is the classical conductive
flux (Spitzer 1962), $\kappa(T) = 9.2\times 10^{-7} T^{5/2}$ erg
s$^{-1}$ K$^{-1}$ cm$^{-1}$ the thermal conductivity, $q_{\rm sat}
=  -\mbox{sign}\left(\nabla T\right)~ 5\phi \rho c_{\rm s}^3$ the
saturated flux, $c_{\rm s}$ the isothermal sound speed. 
The correction factor $\phi=1$ is set according to the values 
suggested for the coronal plasma (Giuliani 1984, Borkowski
\etal 1989, Fadeyev \etal 2002, and references therein).

The flare is triggered by a heat pulse defined by the function 
$Q(s,t) = H_0 \times g(s) \times f(t)$, where $g(s)$ is a Gaussian function:

\[
g(s) = \exp [ - (s - s_H)^2/ (2 \sigma_H ^2)]
\]

\noindent
and $f(t)$ is a pulse function (see Reale \& Orlando 2008 for details). 
In all our simulations, the heat pulse starts at time $t = 0$ and it is 
switched off after 180 s. We do not change this duration, since the late 
evolution of the flare is governed mostly by the total energy input, which 
we tune by changing $H_0$.

The model is implemented and solved numerically using the \FLASH\ code 
(Fryxell \etal 2000), an adaptive mesh refinement multiphysics code, 
extended by additional computational modules to handle the plasma thermal
conduction (see Orlando \etal 2005 for details of the implementation),
the non-equilibrium ionization (NEI) effects (see Reale \& Orlando 2008
for the details), the radiative losses, and the heating function.
 
As for the initial conditions, before the flare the loop plasma is assumed
to be in pressure and energy equilibrium according to Serio \etal (1981) and 
to Orlando \etal (1995). The plasma is very cool and tenuous: the pressure 
at the base of the corona is $p_0 = 0.055$ dyn~cm$^{-2}$, which corresponds 
to a maximum temperature $T_0 \approx 0.8$~MK at the loop apex, according 
to the loop scaling laws (Rosner \etal 1978). Since the pressure increases 
enormously during the flare, the flare evolution is largely independent of 
these initial conditions. The coronal part of the loop is linked to an 
isothermal chromosphere at $T_{\rm c} = 20000$ K and $0.5 \times 10^9$cm 
thick, which acts only as a mass reservoir. All the ion species are assumed 
to be initially in ionization equilibrium everywhere in the loop.

Since we assume that the loop is symmetric with respect to the vertical
axis across the apex, we simulate only half of the loop. For the 
case $L = 3 \times 10^9$ cm, the computational
domain spans a total extension of $3.5 \times 10^9$ cm, including the 
chromosphere which has a thickness of $5 \times 10^8$ cm.
At the coarsest resolution, the adaptive mesh algorithm used in the
\FLASH\ code (\PARAMESH; MacNeice \etal 2000) uniformly covers the
computational domain with a mesh of $16$ blocks, each with $8$ cells.
We allow for 3 levels of refinement, with resolution increasing twice at
each refinement level. The refinement criterion adopted (L\"ohner 1987)
follows the changes in density and temperature. With this grid configuration
the effective resolution is $\approx 3.4\times 10^6$ cm at the
finest level, corresponding to an equivalent uniform mesh of $1024$ grid
points. We use fixed boundary conditions at $s=0$ and reflecting boundary
conditions at $s=s_{\rm max}$ (consistent with the adopted symmetry).

\section{Model results and comparison with observations}
\label{results}

After some exploration of the parameter space, results that provide 
a good match to observations are obtained with a loop of half length 
$L = 3 \times 10^9$ cm, and with a heat pulse of intensity $H_0 = 8$ 
erg~cm$^{-3}$ s$^{-1}$, Gaussian width and center $\sigma_H = 10^8$~cm 
and $s_H = 8 \times 10^8$ cm, respectively. The heat pulse is therefore 
located close the loop footpoint i.e. $\approx 3 \times 10^8$~cm above 
the transition region. We now discuss this simulation in some detail.

\begin{figure}
\includegraphics[width=15.0cm]{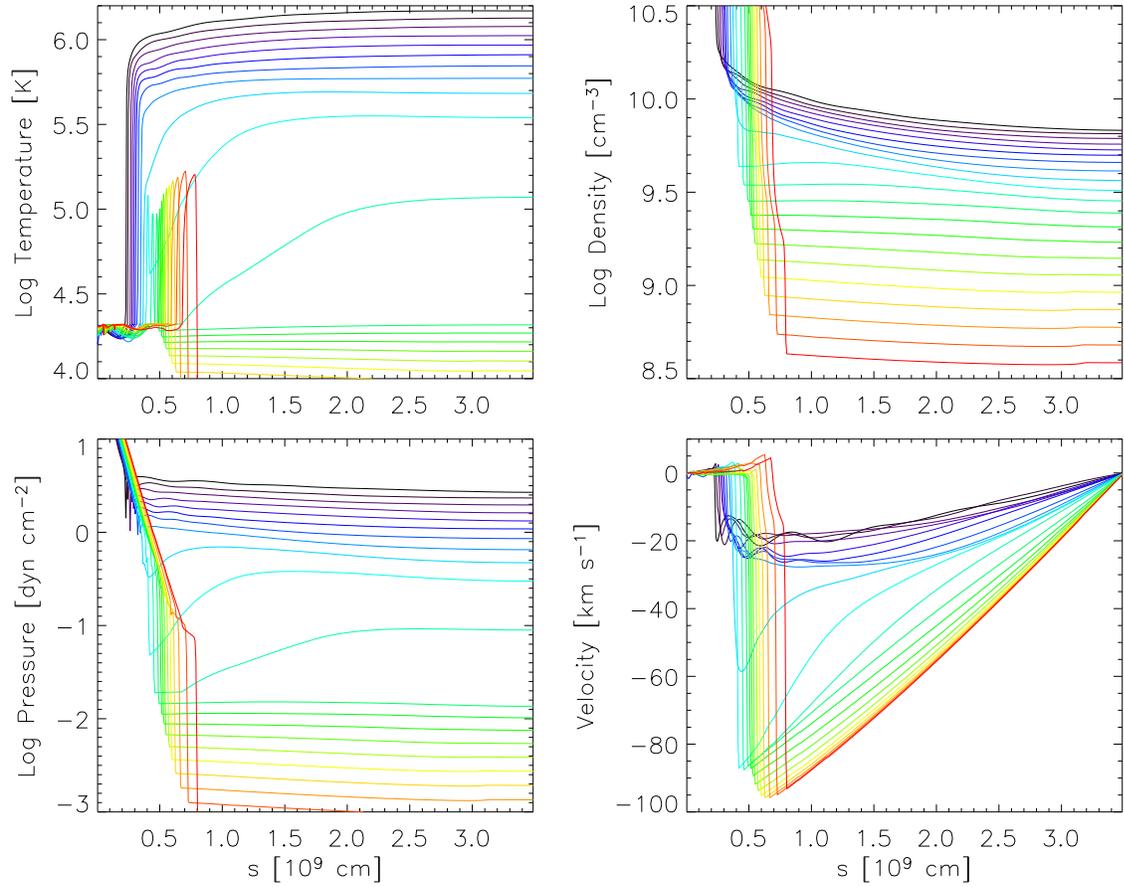}
\caption{Evolution of temperature, density, pressure and velocity along
half of the  model loop in the late phases of the flare, when the plasma
becomes thermally unstable. The curves are sampled at time intervals
of 1 minute between 30 (black line) and 50 (red line) minutes since the
heat pulse has been triggered.}
\label{fig:evol}
\end{figure}

The evolution of the flare-heated plasma confined in a coronal loop
is well known from previous work (e.g. Nagai 1980, Peres et al. 1982,
Cheng et al. 1983, Nagai \& Emslie 1984, Fisher et al. 1985, MacNeice
1986, Betta et al. 2001). In particular, the choice of the parameters
and the specific model make our results very similar to those described
in Reale \& Orlando (2008) for the case of heat pulse deposited at the 
loop footpoints and lasting 180 s. The only difference is that the heating rate
is about 5 times larger, leading to a maximum temperature twice as high,
i.e. about 20 MK.  Since here we focus on the late phases of the flare,
we do not enter the details of the initial phases which are similar
to those described by Reale \& Orlando (2008). We only mention that
the loop is rapidly filled by hot plasma and that the plasma begins to
cool down by conduction and radiation as soon as the heat pulse stops,
but the density still grows for a few minutes. The plasma then drains
and cools altogether. Fig.~\ref{fig:evol} illustrates the evolution of
the plasma along half of the model loop from 30 to 50 minutes after the
start of the heat pulse. During this phase, the plasma density at
the formation temperature of the \ion[Si iii] lines is in the range 
$10 < \log N_{\rm e} < 11$, in agreement with the values inferred from 
spectroscopic analysis of the data (see Sect.~\ref{diagnostics_results}).
The plasma cools down gradually and uniformly along the loop for about 10
minutes but then it suddenly gets thermally unstable and the temperature
drops in about 2 minutes by a factor 10, from $\log T \approx 5.5$
to $\log T \approx 4.5$, in the coronal part of the loop. The density
decreases more gradually and at a constant rate, so that at $t \sim 40$
min the plasma is cold but still relatively dense. Since the plasma is
dramatically out of the hydrostatic equilibrium, it begins to drain more
rapidly and its downward velocity increases from $\sim 20$ km/s to $\sim
100$ km/s. A front of relatively hotter and denser plasma ($\log T \approx 5$) 
develops at the loop footpoints and slowly propagates along the loop.

The trigger of the thermal instability can be explained in terms of 
the criterion described by Field (1965): below 1~MK the radiative losses 
begin to increase for decreasing temperature (negative slope) and the 
density is sufficiently high that the cooling becomes catastrophic.

\begin{figure}
\includegraphics[width=15.0cm]{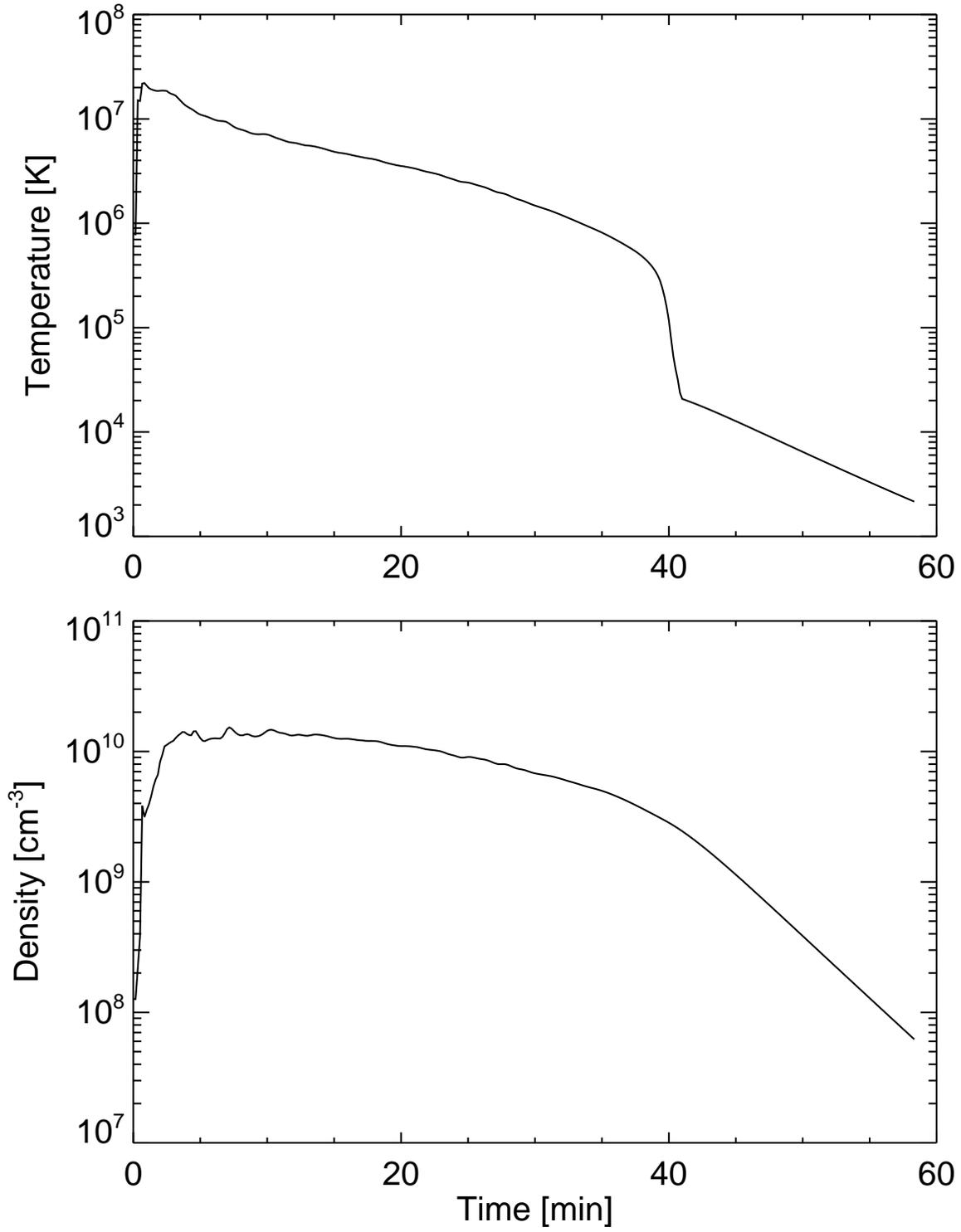}
\caption{Evolution of the temperature (top panel) and density (lower
panel) at the top of the loop throughout the modelled flare duration.}
\label{fig:max_tn}
\end{figure}

The presence of the instability is even more clear in the evolution of the 
loop top temperature. The temperature at the top of the loop is a good 
proxy of the loop maximum temperature for most of the time (Fig.~\ref{fig:evol}). 
Fig.~\ref{fig:max_tn} shows the loop top temperature and density during 
the entire duration of the flare. The heat capacity of the loop is 
initially very small and the temperature increases very rapidly, but it 
also saturates as well because of the increased losses due to the highly 
efficient thermal conduction. After the end of the heat pulse, the
temperature decreases regularly (exponentially) during most of the first 
40 min, and then drops suddenly to $T \sim 20000$ K. Below this temperature, 
the plasma is assumed to radiate no more, so the thermal evolution for 
$t > 41$ min is outside the scope of the modeling. After the temperature 
drops, the plasma begins to drain more rapidly and the 
density also decreases more rapidly (lower panel). 

\begin{figure}
\includegraphics[width=15.0cm]{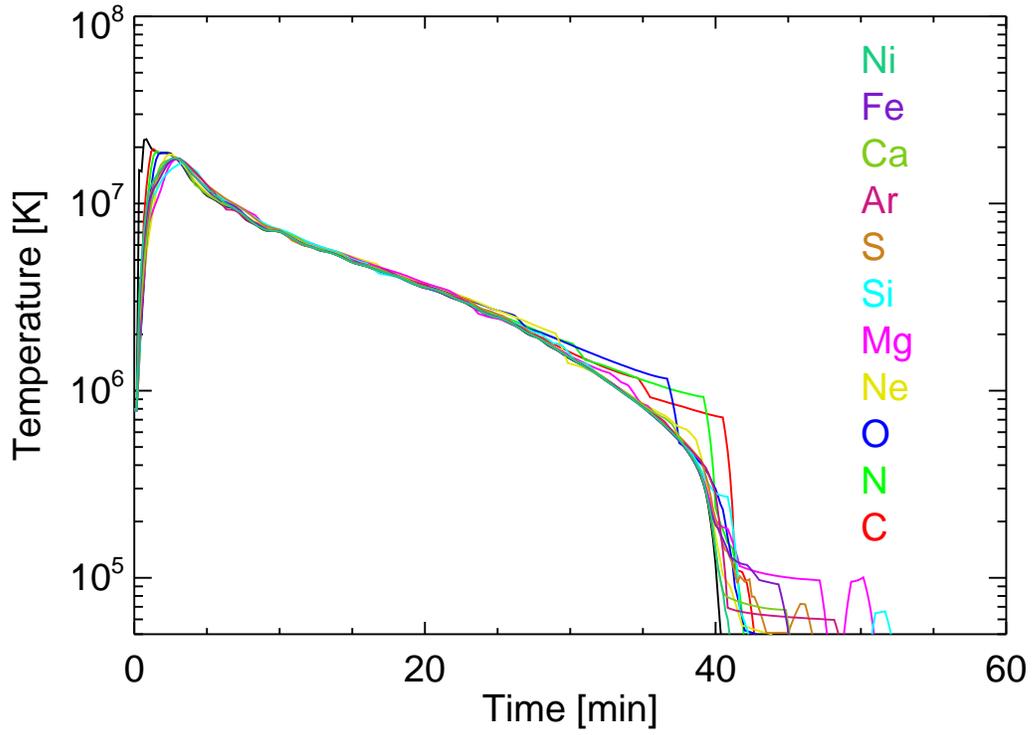}
\caption{Evolution of element effective temperature T$_{eff}$ at the loop top 
during the entire flare evolution. Each color line marks the temperature
related to a different element as derived in Reale \& Orlando (2008) (see text 
for details). The black line indicates the electron temperature.}
\label{fig:max_t}
\end{figure}

To study the degree of departure from equilibrium in a plasma, Reale \&
Orlando (2008) introduced for each element an "effective temperature"
(T$_{eff}$), as the temperature for which the equilibrium ion abundances
of the element best approximates the actual values in the plasma. In
case of equilibrium, all elements have the same T$_{eff}$ and this
should equal the electron temperature. Fig.~\ref{fig:max_t} shows
the evolution of T$_{eff}$ at the loop top for the most important ion
species: the ion temperatures substantially differ from $T_{\rm e}$
only during the earliest phase of the flare ($t< 5$ min) and do not
evolve very differently from $T_{\rm e}$ in later phases. We
find again some deviations from the ionization equilibrium around the
thermal instability ($30 < t < 40$ min), because the temperature drops
abruptly. The instability starts from a hotter status ($\sim 1$ MK) for
the light elements C, N, and O ions (when the electron temperature is
$\approx 5\times 10^5$ K). This temperature drop is mainly due to the
fact that the thermal instability occurs in a range of temperature in
which the He-like ionization state of these light elements is the most
populated. Since the temperature $T_{eff}$ is found as the temperature
at which the two most populated ionization states fit those assuming
equilibrium of ionization, the NEI effects together with the fact that
He-like ions are the most abundant in a broad range of temperatures lead
to the observed jump of $T_{eff}$. Note also that $T_{eff}$ for O, N,
C drop consecutively in time because the corresponding He-like ionization
state forms at lower and lower temperatures during the plasma cooling.
Although apparently large, the deviations from equilibrium are so
concentrated in time that they do not produce strong effect on the
visible emission.

\begin{figure}[htbp]
 \centering
 \includegraphics[width=11.0cm]{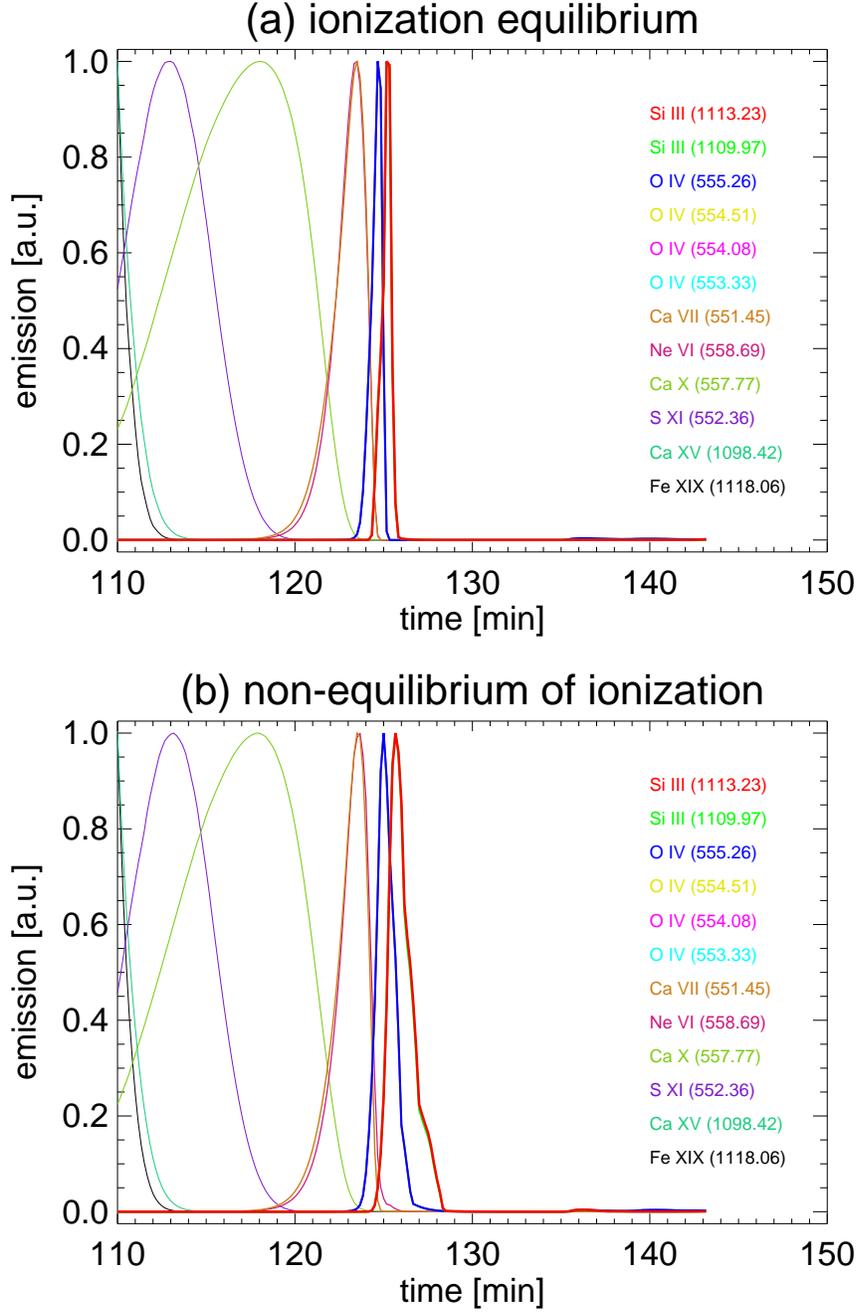}
\caption{Light curves synthesized in the labelled spectral lines from the
flare loop model. The times have been shifted to match the times of the
observation. Only the late phases of the flare are shown, the initial
time (t = 110 min) corresponds to t=25 min since the start of the heat
pulse, and to 20:51 UT in Figure~\ref{light_curves}. (a) Light curves 
computed assuming ionization equilibrium. (b) Light curves computed 
including the effects of non-equilibrium
of ionization.}
\label{fig:lc1l}
\end{figure}

The model allows us to predict the light curves of the observed lines. 
Fig.~\ref{fig:lc1l} shows them during the development of the 
thermal instability in the late phase of the flare, both assuming and not 
assuming ionization equilibrium. Time origin and range have been adjusted to allow
a direct comparison with the observed light curves. Fig.~\ref{fig:lc1l} 
confirms that the effects of deviations from ionization equilibrium are 
not important: 
only the light curves of the coldest lines (\ion[Si iii], \ion[O iv]) are 
affected, by being slightly broader, i.e. their decay is slightly slower 
in the latest phases. Thus, both panels can be compared to observations 
and they show the same results: first, the time sequence and the relative 
start time of the light curves of each line are both in very good agreement with
observations (within 1-2 min), with the exception of the hottest lines 
(\ion[Ca xv]); and second, the observed light curves are considerably 
broader than the predicted ones, i.e. both the rise and the decay are 
slower than modelled. 

We have found that the timing and line sequence could not be reproduced 
with the same accuracy with other model parameters, and in particular longer  
($5 \times 10^9$ cm) loops, more (to a maximum temperature of 25~MK) or less 
(10~MK) intense heating, and also a heat pulse deposited at the loop apex. 
We define the root mean square time distance of the simulated and observed 
peak between lines as a goodness-of-fit parameter:

\[
\sigma_t = \sum_i \sqrt{ \left( t_{sim} - t_{obs} \right)^2_i}
\] 

where $t_{sim}$ and $t_{obs}$ are the times of the peak of the i-th line light 
curves measured in the simulation and in the observation, respectively. Here we 
align the times to be the same for the peak of the \ion[Si iii] line.
We measure $\sigma_ t$ over the coolest lines (from S xi to Si iii).

We have found that the value of the best-fit model is $\sigma_t \sim 4$
min, while for all others values between 9 and 50 min. For the best-fit
model, each peak of the (cool) model light curves is within 5 min from
respective peak of the observed light curves. Indeed, we have found
$\sigma_t \sim 4$ min also from a simulation with a shorter loop $L = 1.5
\times 10^9$~cm and maximum temperature of 20 MK.

One more constraint from the observation is the absolute intensity of the 
line emission. To compute the line intensity from the simulation results we 
need to make an assumption on the path length $H$ along the line of sight. For 
$L = 3 \times 10^9$ cm, we find that the path length that matches the measured 
one is between $1$ and $2 \times 10^8$ cm for the lines marking the thermal 
instability (from the warm \ion[S xi] to the cool \ion[Si iii]). This is 
in agreement with the estimate of $H$ made in Section~\ref{diagnostics_results}, 
and it is also compatible with a loop cross-section diameter of the order of 
10\% of the loop length, as typical of coronal loops. For the shorter loop 
($L = 1.5 \times 10^9$ cm) the path length is larger ($1.5 - 4 \times 10^{8}$ cm)
but still well within the range of values found in Section~\ref{diagnostics_results}. 
In this case, the flaring structure would have more the aspect of a flaring loop 
arcade. Since we have reasons to believe that the flare occurs in a multi-structured 
loops, both scenarios seem possibile. Morevover, these loop lengths are compatible 
with the vertical height corresponding to the SUMER slit position above the limb.
Overall, we can say that the loop length, and the heat 
intensity and position are relatively well constrained and provide a reasonably 
self-consistent scenario. 

None of the explored single loop models has been able to reproduce the observed 
"broad" light curves, i.e. the observed light curves change more slowly 
than those obtained from the hydrodynamic simulations.

\begin{figure}[htbp]
 \centering
   {\includegraphics[width=11cm]{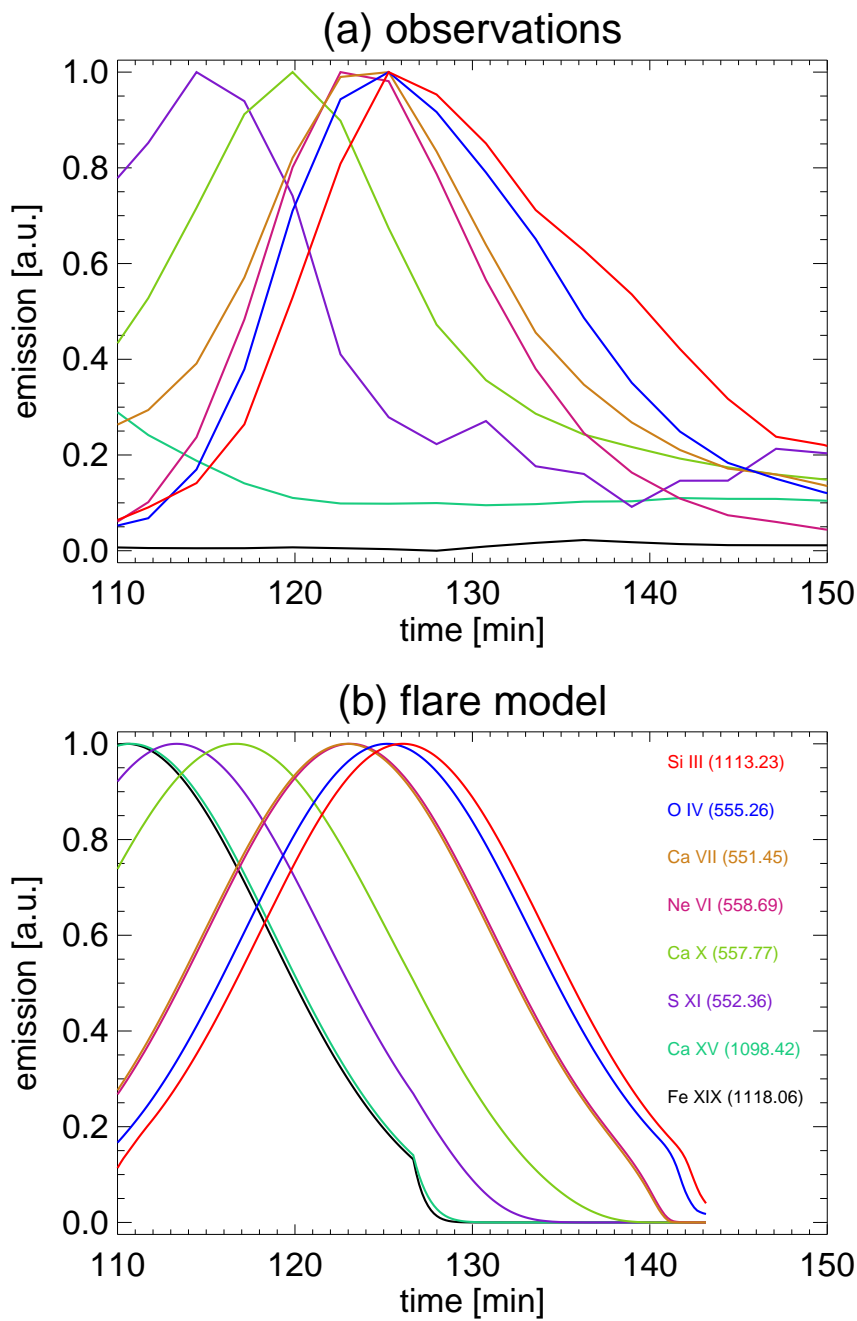}}
\caption{Comparison of (a) observed light curves with (b) best model light
curves obtained assuming a distribution of heat pulses in different loop structures. 
In particular, curves in (b) are obtained convolving those in Fig.~\ref{fig:lc1l}b 
with a Gaussian whose width is 8 min. The initial time (t = 110 min) corresponds to 
t=25 min since the start of the heat pulse, and to 20:51 UT in Figure~\ref{light_curves}.}
\label{fig:lcok}
\end{figure}

The agreement is further improved if we assume that the flaring region is 
not a single monolithic loop, but it is substructured into a bundle of 
loops or loop strands, each of the same length as the loop we modeled. Each 
component is heated at a different time, and the time distribution of the 
heating pulses is assumed to  be Gaussian with $\sigma=8$~minute width. We 
derive the related line light curves simply as the convolution of the single 
loop light curves with the (same) Gaussian. The results are shown in 
Fig.~\ref{fig:lcok} and compared to observations. The predicted light 
curves are now in closer agreement as far as time sequence, time 
evolution and time scale are concerned. 

We are unable to, and do not pretend to, constrain the number of loops or 
strands. The convolution with the Gaussian means that at any time the line 
intensities are the same relative to the others (i.e. the ratio is unchanged). 
So there is only a scale factor that changes in time, which is a volume factor. 
This  implies that all loops/strands are heated in the same way, and only the 
number of heated loops/strands first progressively increases and then 
progressively decreases. We cannot claim that this is the only way to 
produce this effect, but we can certainly say that this is a very simple 
way to do that, with very few free parameters (only the rate of involved 
strands).

Although fine details such as the asymmetry between rise and decay of the 
light curves can still be improved, overall we can conclude that this 
agreement is evidence of a thermally unstable plasma confined in a 
substructured loop system undergoing cooling after a flare.

It is worth noting that the thermal instability occurs at $t \sim 40$ 
min after the heat pulse, while the observation indicates that it should 
occur $\sim 60$ min after the start of the flare (see Fig.2). 
The discrepancy between the flare start time and the start of the heat 
pulse, as well as the disagreement in the time profile of hot lines, may be 
better resolved if the model can produce a temperature time profile that at 
first decreases from 20~MK more rapidly but stays longer (by $\simeq$20 min) 
above $\simeq$1~MK before going into thermal instability. This appears to be 
limited by the current setting of the model for the energy balance.


\section{Discussion and Conclusions}
\label{discussion}

This work aims at explaining the peculiar light curves of several UV lines
in the late phases of a flare observed with SOHO/SUMER. In particular, our
target is the appearance of several spectral lines emitted by plasma
in a wide range of temperature in a relatively short time. The scope of 
this work is modeling the post-flare phase and therefore we do not attempt 
at reproducing the initial phases of the flare. 

We show that light curves are well reproduced by the hydrodynamic evolution 
of flaring plasma confined in a closed loop system triggered by a strong 
heat pulse. In particular, the sudden, almost simultaneous appearance of 
the emission of chromospheric and transition region ions formed more than 
one order of magnitude apart in temperature is due to the fact that late 
in the cooling phase the plasma becomes thermally unstable, because 
the radiative losses increase rapidly below 1 MK, and its temperature 
drops rapidly by one order of magnitude in a couple of minutes along most 
of the loop. A good relative timing and order of appearance of the light 
curves of each line is obtained with a standard flare loop model considering 
a loop half-length of $\leq 3 \times 10^9$ cm, and a heat pulse lasting 
3~min and deposited in the corona, close to the loop footpoints. The 
heating rate is such as to have a flare maximum temperature of about 20~MK. 

Better agreement of the duration of the light curve of each line is 
obtained if we assume that the flare occurs in a bundle of subloops of 
equal length (a stranded loop or a loop arcade). If each substructure is 
assumed to be heated at a different time, the predicted light curves can 
be lengthened to match the observed ones. This is consistent with the slit 
intercepting several concentric arches that undergo independently the same 
evolution, so that we see the envelope of this close sequence of events. 
The time scale of the distribution of the events is about 8 min, that indicates 
a propagation speed of the trigger signal of $\sim 10$ km/s over a length scale 
of $\sim 10^9$ cm. This result might be consistent with the presence of fine heat 
substructuring during flares (e.g. Antiochos \& Krall 1979, Reeves \& Warren 
2002, Warren 2006).

The occurrence of the thermal instability is a necessary ingredient to 
explain the observations. However, the effects of such instability on the 
ion fractions of the major elements are rather limited, contrarily to our
initial expectations. The reason of such limited departures is the relatively
large density of the cooling plasma, which shortens the ionization and
recombination times scales enough to allow the plasma to efficiently adjust
its ionization status to the rapidly decreasing electron temperature (e.g. 
Golub \etal 1989). 

We do not reproduce all the details of the observations. In 
particular, the light curves of the hot lines are not well reproduced, 
and probably would need a more complete analysis of the entire flare 
event, which is not the target of this study.

The main limitation of the present work is the lack of spatial information
on the flaring region, and better constraints on the heating location
and timing, and on the properties of the flaring loops systems, can
be obtained by combining the present modeling with flares observed with
the EVE spectrometer and the AIA imaging system on the SDO.

\acknowledgements

The work of Enrico Landi is supported by the NNX10AM17G and NNX11AC20G
NASA grants. Fabio
Reale and Salvatore Orlando acknowledge support from Italian Ministero
dell'Universit\'a e Ricerca and from Agenzia Spaziale Italiana (ASI),
contract I/015/07/0. The software used in this work was in part developed
by the DOE-supported ASC / Alliance Center for Astrophysical Thermonuclear
Flashes at the University of Chicago. The simulations have been executed
at the HPC facility (SCAN) of the INAF-Osservatorio Astronomico di
Palermo. The authors wish to thank the anonymous referee for his/her
comments that helped us improve significantly the paper.

\begin{table}
\begin{center}
\begin{tabular}{lrcl}
Ion & Wavelength (\AA) & $\log T_{max}$ & Transition \\
\hline
 & \\
\ion[Si iii]   & 1109.97 & 4.7 & \orb[3s3p ] \tm[3 P 1] - \orb[3s3d ] \tm[3 D 1,2] \\
\ion[Si iii]   & 1113.23 & 4.7 & \orb[3s3p ] \tm[3 P 2] - \orb[3s3d ] \tm[3 D 1,2,3] \\
\ion[O iv]     &  553.33 & 5.2 & \orb[2s 2]\orb[2p ] \tm[2 P 1/2] - \orb[2s2p 2] \tm[2 P 3/2] \\
\ion[O iv]     &  554.08 & 5.2 & \orb[2s 2]\orb[2p ] \tm[2 P 1/2] - \orb[2s2p 2] \tm[2 P 1/2] \\
\ion[O iv]     &  554.51 & 5.2 & \orb[2s 2]\orb[2p ] \tm[2 P 3/2] - \orb[2s2p 2] \tm[2 P 3/2] \\
\ion[O iv]     &  555.26 & 5.2 & \orb[2s 2]\orb[2p ] \tm[2 P 3/2] - \orb[2s2p 2] \tm[2 P 1/2] \\
\ion[Ne vi]    &  558.69 & 5.6 & \orb[2s 2]\orb[2p ] \tm[2 P 1/2] - \orb[2s2p 2] \tm[2 D 3/2] \\
\ion[Ca vii]   &  551.45 & 5.7 & \orb[3s 2]\orb[3p 2] \tm[3 P 2] - \orb[3s3p 3] \tm[3 P 1,2] \\
\ion[Ca x]     &  557.77 & 5.9 & \orb[3s ] \tm[2 S 1/2] - \orb[3p ] \tm[2 P 3/2] \\
\ion[Al xi]    &  550.03 & 6.1 & \orb[2s ] \tm[2 S 1/2] - \orb[2p ] \tm[2 P 3/2] \\
\ion[S xi]     &  552.36 & 6.3 & \orb[2s 2]\orb[2p 2] \tm[3 P 1] - \orb[2s2p 3] \tm[5 S 2] \\
\ion[Na x]     & 1111.77 & 6.3 & \orb[1s2s ] \tm[3 S 1] - \orb[1s2p ] \tm[3 P 2] \\
\ion[Ca xv]    & 1098.42 & 6.6 & \orb[2s 2]\orb[2p 2] \tm[3 P 1] - \orb[2s 2]\orb[2p 2] \tm[1 D 2] \\
\ion[Fe xix]   & 1118.06 & 7.0 & \orb[2s 2]\orb[2p 4] \tm[3 P 2] - \orb[2s 2]\orb[2p 4] \tm[3 P 1] \\
 & \\
\hline
\end{tabular}
\end{center}
\caption{Lines included in the SUMER spectra. $T_{max}$ is the temperature
of maximum abundance of the ion, from Bryans \etal (2009). Lines with 
wavelengths smaller than 1000~\AA\ were observed in second order.}
\label{lines}
\end{table}

\end{document}